\documentclass[aps,prb,twocolumn,groupedaddress,floatfix,showpacs]{revtex4}

\usepackage{subfigure}
\usepackage{color}
\usepackage{graphicx}
\usepackage{amssymb}
\usepackage{amsmath}
\usepackage{natbib}

\begin{document}

\newcommand{\eps}{\varepsilon}
\newcommand{\ud}{\mathrm{d}}
\newcommand{\pd}{\partial}
\newcommand{\dx}{\partial_x}
\newcommand{\dt}{\partial_t}
\newcommand{\vd}{\delta}
\newcommand{\Tr}{\mathrm{Tr}}
\newcommand{\T}{\mathrm{T}}
\newcommand{\bq}{\begin{equation}}
\newcommand{\eq}{\end{equation}}
\newcommand{\bea}{\begin{eqnarray}}
\newcommand{\eea}{\end{eqnarray}}
\newcommand{\bef}{\begin{figure}}
\newcommand{\ef}{\end{figure}}
\newcommand{\bdm}{\begin{displaymath}}
\newcommand{\edm}{\end{displaymath}}
\newcommand{\ba}{\begin{array}}
\newcommand{\ea}{\end{array}}

\newcommand{\kk}{\mathbf{k}}
\newcommand{\QQ}{\mathbf{Q}}
\newcommand{\qq}{\mathbf{q}}
\newcommand{\pp}{\mathbf{p}}
\newcommand{\up}{\uparrow}
\newcommand{\dn}{\downarrow}
\newcommand{\spin}{\mathbf{S}}
\newcommand{\om}{\omega}

\newcommand{\uu}{\up\!\up}
\newcommand{\updn}{\up\!\dn}
\newcommand{\dnup}{\dn\!\up}
\newcommand{\dd}{\dn\!\dn}
\newcommand{\upd}{\up\dn}

\newcommand{\Tuu}{T_{\uu}}
\newcommand{\Tdd}{T_{\dd}}
\newcommand{\Tud}{T_{\updn}}
\newcommand{\Tdu}{T_{\dnup}}

\newcommand{\g}{\gamma}
\newcommand{\D}{\Delta}
\newcommand{\la}{\langle}
\newcommand{\ra}{\rangle}
\newcommand{\wn}{{\omega_n}}

\newcommand{\xx}{\mathbf{x}}
\newcommand{\ee}{\mathbf{e}}

\newcommand{\kr}{k_R}
\newcommand{\psirup}{\psi_{R\up}}
\newcommand{\psirdn}{\psi_{R\dn}}
\newcommand{\psilup}{\psi_{L\up}}
\newcommand{\psildn}{\psi_{L\dn}}
\newcommand{\psirs}{\psi_{R\sigma}}
\newcommand{\psils}{\psi_{L\sigma}}

\newcommand{\rup}{{R\up}}
\newcommand{\rdn}{{R\dn}}
\newcommand{\lup}{{L\up}}
\newcommand{\ldn}{{L\dn}}

\newcommand{\gperp}{g_{2\bot}}
\newcommand{\gpar}{g_{2\|}}
\newcommand{\gfperp}{g_{4\bot}}
\newcommand{\gfpar}{g_{4\|}}
\newcommand{\udx}{\ud x}
\newcommand{\norm}{\boldsymbol{:}}

\newcommand{\tv}{\tilde{v}}
\newcommand{\tK}{\tilde{K}}
\newcommand{\tphi}{\tilde{\phi}}
\newcommand{\ttheta}{\tilde{\theta}}
\newcommand{\tG}{\tilde{G}}

\parskip=0pt

\title{Magnetism and d-wave superconductivity on the half-filled square lattice with frustration}
{\rm }


\author{Andriy H. Nevidomskyy$^1$}
\email[\vspace{-1mm} E-mail: ]{nevidomskyy@cantab.net}
\altaffiliation{Present address: Department of Physics and Astronomy,
  Rutgers University, 136 Frelinghuysen Road, Piscataway, NJ 08854, USA}


\author{Christian Scheiber$^2$}

\author{David S\'en\'echal$^1$}
\author{A.-M. S. Tremblay$^1$}
\affiliation{$^1$D\'epartement de physique, Universit\'e de Sherbrooke,
  Sherbrooke, Qu\'ebec, J1K 2R1, Canada}
\noaffiliation
\affiliation{$^2$Institute of Theoretical and Computational Physics,
Graz University of Technology, Petersgasse 16, 8010 Graz, Austria}
\noaffiliation



\date{\today}

\begin{abstract}
The role of frustration and interaction strength on the half-filled Hubbard model is studied on the square
lattice with nearest and next-nearest neighbour hoppings $t$ and $t'$ using
the Variational Cluster Approximation (VCA). At half-filling, we find
two phases with long-range antiferromagnetic (AF) order: the usual
N\'eel phase, stable at small frustration $t'/t$, 
and the so-called collinear (or super-antiferromagnet) phase with ordering
wave-vector $(\pi,0)$ or $(0,\pi)$, stable for large frustration.
These are separated by a phase with no detectable long-range magnetic order.
We also find the d-wave superconducting (SC) phase ($d_{x^2-y^2}$),
which is favoured by frustration if it is not too large. 
Intriguingly, there is a broad region of coexistence where both AF and
SC order parameters have non-zero values.
In addition, the physics of the metal-insulator transition in the normal state is analyzed. 
The results obtained with the help of the VCA method are compared
with the large-$U$ expansion of the Hubbard model and known results
for the frustrated $J_1$--$J_2$ Heisenberg model. These results are
relevant for pressure studies of undoped parents of the high-temperature
superconductors: we predict that an insulator to d-wave SC
transition may appear under pressure. 
\end{abstract}



\maketitle

\section{Introduction}
The subject of frustration~\cite{Moessner01} in quantum magnetic systems has received
increased attention in recent years, fuelled in part by the discovery~\cite{Bednorz86}
of high-temperature superconductivity in the doped cuprates.
It is believed to play a key role in a number of recently observed
phenomena, such as the large anomalous Hall effect in ferromagnetic
pyrochlores~\cite{pyrochlores_Taguchi01}, the unconventional
superconductivity in water substituted sodium cobaltate Na$_x$CoO$_2$,
which is composed of triangular sheets of Co
atoms~\cite{cobaltates_Takada03}, the interplay between magnetism and
unconventional superconductivity in organic layered compounds of the
$\kappa$-BEDT family~\cite{BEDT_Cl,BEDT_CN3} or
the interaction between electric and magnetic properties in
multiferroic materials~\cite{multiferroics_Blake05}.

The issue of frustration has been studied mostly on
two classes of theoretical models: spin Hamiltonians, such as the
$J_1$-$J_2$ Heisenberg model discussed below, and toy dimer models, the latter
inspired by P.W.~Anderson's proposal~\cite{Anderson87} of the resonating valence-bond (RVB)
state as a possible explanation for high-$T_c$ superconductivity.

One can view spin Hamiltonians as the large interaction, $U$, limit of the Hubbard model. It is thus of interest to study the effect of both interaction and frustration on the phase diagram. In this work, we  study systematically the frustrated
Hubbard model at half-filling on a square lattice with nearest $t$ and next-nearest
neighbour $t'$ hoppings, described by the Hamiltonian:
\bq
H = t\sum_{<i,j>}c_i^\dagger c_j +  t'\sum_{\ll i,j\gg}c_i^\dagger c_j
+ U\sum_i n_{i\up} n_{i\dn},
\label{Hubbard}
\eq
where $c_i^\dagger$, $c_i$ are the electron creation and annihilation
operators and $n_{i\sigma}$ is the particle number operator on
site~$i$. The interaction is represented by $U$. Next-nearest neighbor
hopping $t'$ introduces frustration since, from a weak coupling point
of view, it produces deviations from perfect nesting, and, from a
strong-coupling point of view, it leads to an effective
antiferromagnetic superexchange interaction $J_2$ that opposes the
tendency of next-nearest neighbors to order ferromagnetically when
nearest-neighbor superexchange $J_1$ is antiferromagnetic.

Our study of the phase diagram as a function of $U/t$ and $t'/t$ can
also be understood as a study of the generalized zero-temperature
phase diagram for high-temperature superconductors illustrated in
Fig.~\ref{Perspective}. The thin parallelepiped represents
schematically the region of parameter space where families of
high-temperature superconductors appear. We are studying the
zero-doping plane $\delta=0$, where one normally encounters the
insulating antiferromagnetic parents of high-temperature
superconductors. We will see that d-wave superconductivity can also
occur in this plane, so high-pressure studies might conceivably lead
to the observation of d-wave superconductivity even at half-filling,
provided the on-site interaction $U/t$ is not too large.
 The generalized phase diagram also leads to insights into the nature
 of d-wave superconductivity, as we will see. 

We use a quantum cluster approach, the so-called variational cluster
approximation (VCA, sometimes referred to as the 'variational cluster
perturbation theory')~\cite{VCA}. This method has already been used
successfully for the high-temperature
superconductors~\cite{Senechal:2005,Aichhorn:2005,Aichhorn:2007}.
Other quantum cluster methods that are extensions of dynamical
mean-field theory (DMFT)~\cite{DMFT}, such as cellular dynamical
mean-field theory~\cite{CDMFT} and dynamical cluster
approximation~\cite{DCA}, have yielded comparable
results~\cite{Kancharla:2005,Maier:2005} for the same problem. On the
anisotropic triangular lattice at half-filling, both
VCA~\cite{Sahebsara06} and CDMFT~\cite{Kyung06} give a phase diagram
that is in remarkable agreement with that of layered BEDT organic
superconductors.

Wherever possible, we will make connections with earlier numerical studies on the half-filled square-lattice Hubbard model~\cite{Yokoyama06, Mizusaki06}, and with work on spin-Hamiltonians.

\bef[tbp]
\begin{center}{\includegraphics[angle=-90,width=8.6cm]{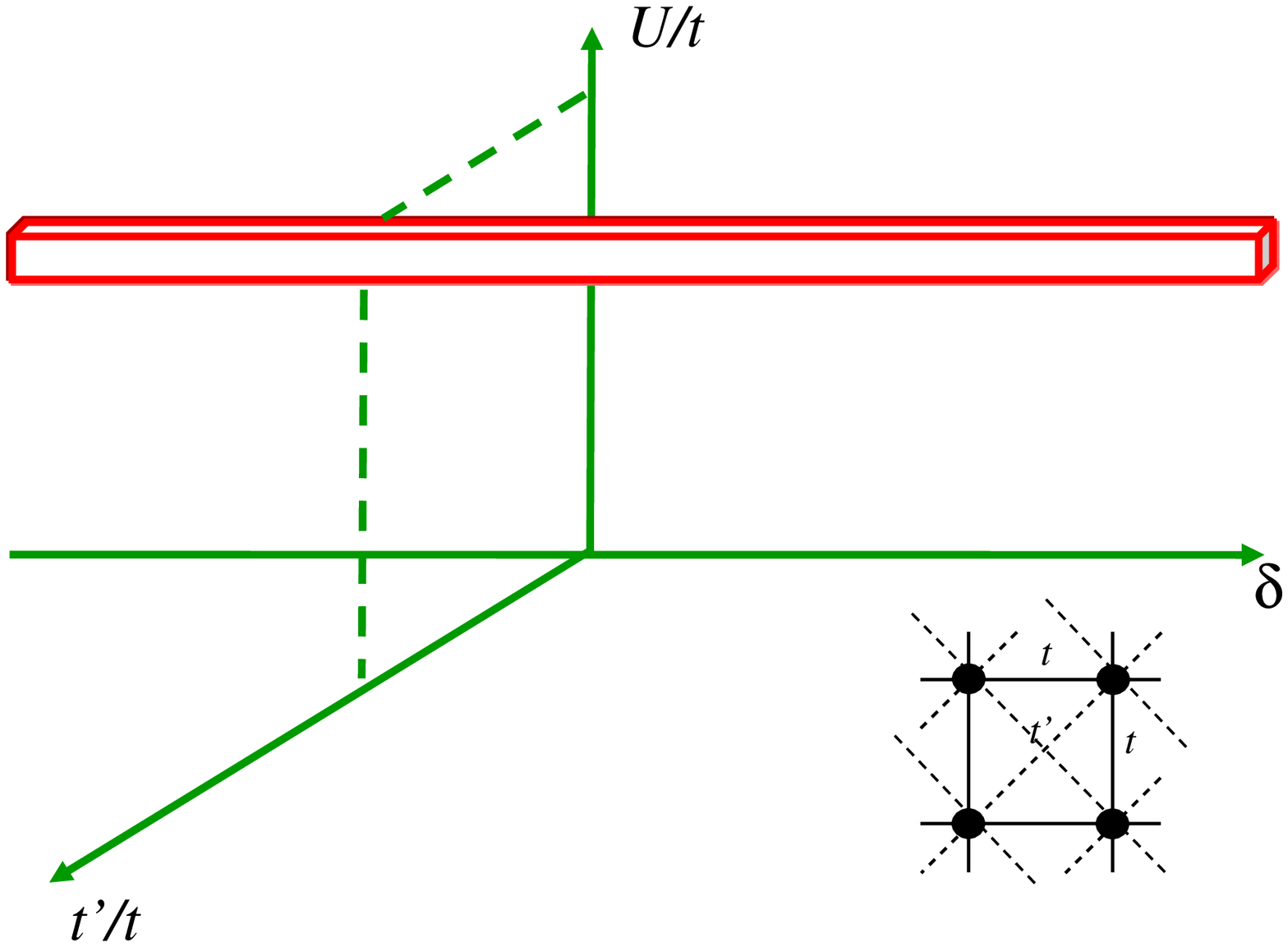}}
\end{center}
\caption
{(Color online) Schematic generalized zero-temperature parameter space
  for the high-temperature superconductors. Horizontal axis $\delta$
  represents doping, vertical axis $U/t$ interaction strength, and
  third direction $t'/t$ frustration. 
The parallelepiped indicates the region of parameter space relevant
for high-temperature superconductors. In this study, we consider the
zero doping $\delta=0$ plane. Experimentally, $U/t$ can be varied by
pressure. The inset shows the definitions of $t$ and $t'$ on the
square lattice. 
}
\label{Perspective}
\ef

This paper is organized as follows. First, the role of frustration
in quantum magnets is illustrated in Section~\ref{Sec.J1-J2} on the well-studied example of
the $J_1$-$J_2$ Heisenberg spin model. The connection to the Hubbard
model is then made by virtue of the large-$U$ expansion in
Section~\ref{Sec.large-U}.
The framework
of the variational cluster approximation (VCA) used in this work is
briefly described in Section~\ref{Sec.VCA}.
The resulting magnetic phase diagram of the Hubbard model is then presented in
Section~\ref{Sec.results}, with a separate subsection devoted to the analysis of the
metal-insulator transition. The main result of this work, where d-wave
superconductivity and magnetism compete and even coexist, is described
in  Section~\ref{Sec.AF-SC}. 
We conclude by discussing the obtained phase
diagram of the frustrated Hubbard model and draw comparison with other
known results in Section~\ref{Sec.conclusions}.

\section{A reference point: \boldmath{$J_1$-$J_2$} Heisenberg model on the
  square lattice}\label{Sec.J1-J2}

One of the earliest studied models that exhibits
frustration is the so-called $J_1$-$J_2$ Heisenberg model, which
contains antiferromagnetic (AF) spin-spin interaction between
nearest and next-nearest neighbours (denoted by $<\!i,j\!>$ and $\ll\!
i,j\!\gg$ respectively):
\bq
H = J_1\sum_{<i,j>}\mathbf{S}_i\cdot \mathbf{S}_j + J_2\sum_{\ll i,k\gg}\mathbf{S}_i\cdot \mathbf{S}_k
\eq
Albeit simple in appearance, this model captures a  number of
important features common to a large class of frustrated quantum
magnets.

Classically, the ground state of the model can be derived by considering the Fourier transform of the spin coupling $J(q)$,
which on the square lattice with next-nearest neighbour spin
interaction takes the following form:
\bq
J(q) = 2J_1(\cos q_x + \cos q_y) + 4J_2 \cos q_x\cos q_y.
\eq
The classical ground state should minimize this coupling, leading to
two possible solutions: the N\'eel state (referred to as AF1 in the following) with the ordering wave-vector
$\mathbf{Q}=(\pi,\pi)$ for the range of parameters $J_2/J_1 < 0.5$,
 and the so-called super-antiferromagnetic phase with $\mathbf{Q}=(\pi,0)$ or
 $(0,\pi)$ (referred to as AF2), realised for  $J_2/J_1 > 0.5$.
Although non-collinear spin states with the same classical ground
state energy can also be realized, it has been
shown~\cite{Shender} that thermal or quantum fluctuations will favour
the states that have collinear magnetization.

The effect of quantum fluctuations becomes especially important around
the quantum critical point $J_2/J_1=0.5$ where the classical ground
state is highly degenerate. The large-$S$ analysis
shows~\cite{Chandra88} that even to the lowest order in $1/S$,
zero-temperature quantum corrections to the sublattice magnetization
diverge at the critical point, pointing to the existence of a quantum
disordered phase.
The nature of such a phase can be captured by dimer covering of the
lattice, which is a caricature for the singlet pairings (i.e. valence bonds) of
nearest-neighbour spins.

A wide literature~\cite{review_VBS} exists on the subject of spin rotationally
invariant dimer order in frustrated quantum magnets. It is generally
believed~\cite{review_VBS} that in the case of a square lattice, the dimer phase
exhibits long-range order in the dimer-dimer correlation functions,
leading to the notion of the `valence bond solid' (VBS), as opposed to
the original RVB phase of Anderson~\cite{Anderson87} which is supposed
to have only short-range order and gapped collective excitations.
We shall touch upon this subject in Section~\ref{Sec.results}, although
this study will be
primarily concerned with the magnetic broken-symmetry phases.

\section{Large-\boldmath{$U$} expansion of the $\boldmath{t-t'-U}$
  Hubbard model}
\label{Sec.large-U}

\bef[!tbp]
\begin{center}{\includegraphics[width=8.6cm]{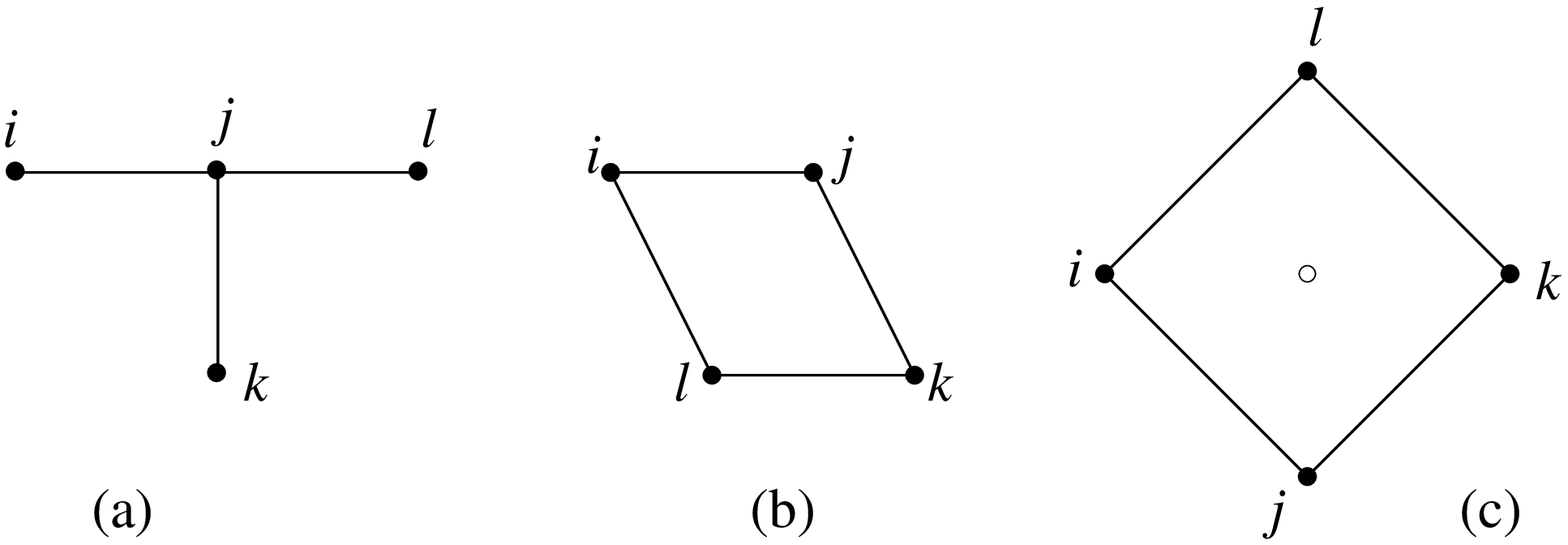}}
\end{center}
\caption
{The ring exchange contributions to the large-$U$ expansion of the
  $t-t'-U$ Hubbard model from Ref.~\onlinecite{Delannoy}. The empty circle
  in case (c) denotes that the central site does not participate in
  the plaquette spin-exchange term.
}
\label{Fig.rings}
\ef

In order to get insight into the physics of the frustrated Hubbard
model, we shall first consider its large-$U$ expansion. Whereas the
procedure for obtaining the low-energy Heisenberg Hamiltonian from the
conventional Hubbard model with nearest-neighbour interaction
is a textbook example~\cite{Auerbach}, the presence of
next-nearest neighbour terms and calculation to order $1/U^3$ leads
non-trivial next-nearest-neighbor and ring-exchange terms.
We exploit here the coefficients of the expansion that have been obtained by Delannoy \emph{et
  al.}~\cite{Delannoy, Delannoy05} by means of the canonical transformation approach. The resulting
effective spin Hamiltonian can be written as follows:

\bea
H &=& J_1\sum_{<i,j>}\mathbf{S}_i\cdot \mathbf{S}_j + J_2\sum_{\ll
  i,k\gg}\mathbf{S}_i\cdot \mathbf{S}_k 
\nonumber\\
  &+&\;\{\mbox{ring exchange terms}\}
\label{H}
\eea
where the coefficients $J_i$ are given by
\bea
J_1 &=& \frac{4t^2}{U} - \frac{24t^4}{U^3}+4\frac{t^2
  t'^2}{U^3}+\dots;\nonumber\\
J_2 &=& \frac{4t^4}{U^3}+\frac{4t'^2}{U}-8\frac{t^2
  t'^2}{U^3} + \dots
\eea
The relevant ring exchange terms~\cite{Delannoy} are defined on the plaquettes
depicted in Fig.~\ref{Fig.rings} with the corresponding analytical expressions given
by:
\bea
H_{\mathrm{(a)}} = 20J_{tt'}P_1(i,j,k,l) - J_{tt'}P_2(i,j,k,l)\nonumber\\
H_{\mathrm{(b)}} = 20J_{tt'}P_1(i,j,k,l) +
J_{tt'}P_2(i,j,k,l)\\
H_{\mathrm{(c)}} = 80\frac{t'^4}{U^3}P_1(i,j,k,l) - J_{tt'}P_2(i,j,k,l)\nonumber
\label{ring_exch}
\eea
where $J_{tt'}=4t^2 t'^2/U^3$ and the following notations have been
used following Ref.~\onlinecite{Delannoy}:
\bea
P_1(i,j,k,l) &=& (\spin_i\cdot\spin_j)(\spin_k\cdot\spin_l) +(\spin_i\cdot\spin_l)(\spin_k\cdot\spin_j)\nonumber\\
&-&(\spin_i\cdot\spin_k)(\spin_j\cdot\spin_l) \\
P_2(i,j,k,l)  &=&  \left\{ \spin_i\cdot\spin_j + \spin_i\cdot\spin_k +
\spin_i\cdot\spin_l\right. \nonumber\\
&+& \left.\spin_j\cdot\spin_k + \spin_j\cdot\spin_l + \spin_k\cdot\spin_l\right\}
\eea

Evaluating the classical ground state energies of the Hamiltonian
Eqs.~(\ref{H}-\ref{ring_exch}) that corresponds to the two possible
ordering wave-vectors $\QQ_1=(\pi,\pi)$ and $\QQ_2=(\pi,0)$, yields
the following result:
\bea
E_{(\pi,\pi)} - E_{(\pi,0)} = \frac{2t^2}{U}\left(-1 +
\frac{8t^2}{U^2}- \frac{12t'^2}{U^2}+2\frac{t'^2}{t^2}\right)
\label{deltaE}
\eea
The corresponding classical phase diagram that follows from this
is shown in Fig.~\ref{Fig.classical}. Note that in the large
of $U$ limit, it follows from Eq.~(\ref{deltaE}) that the criterion
for the $(\pi,0)$ phase to have lower ground state energy is given by
$t'/t>1/\sqrt{2}$, i.e. $J_2/J_1>0.5$, which coincides with the
classical criterion obtained earlier for
the $J_1$-$J_2$ Heisenberg model.

\bef[!tbp]
\begin{center}{\includegraphics[width=8.6cm]{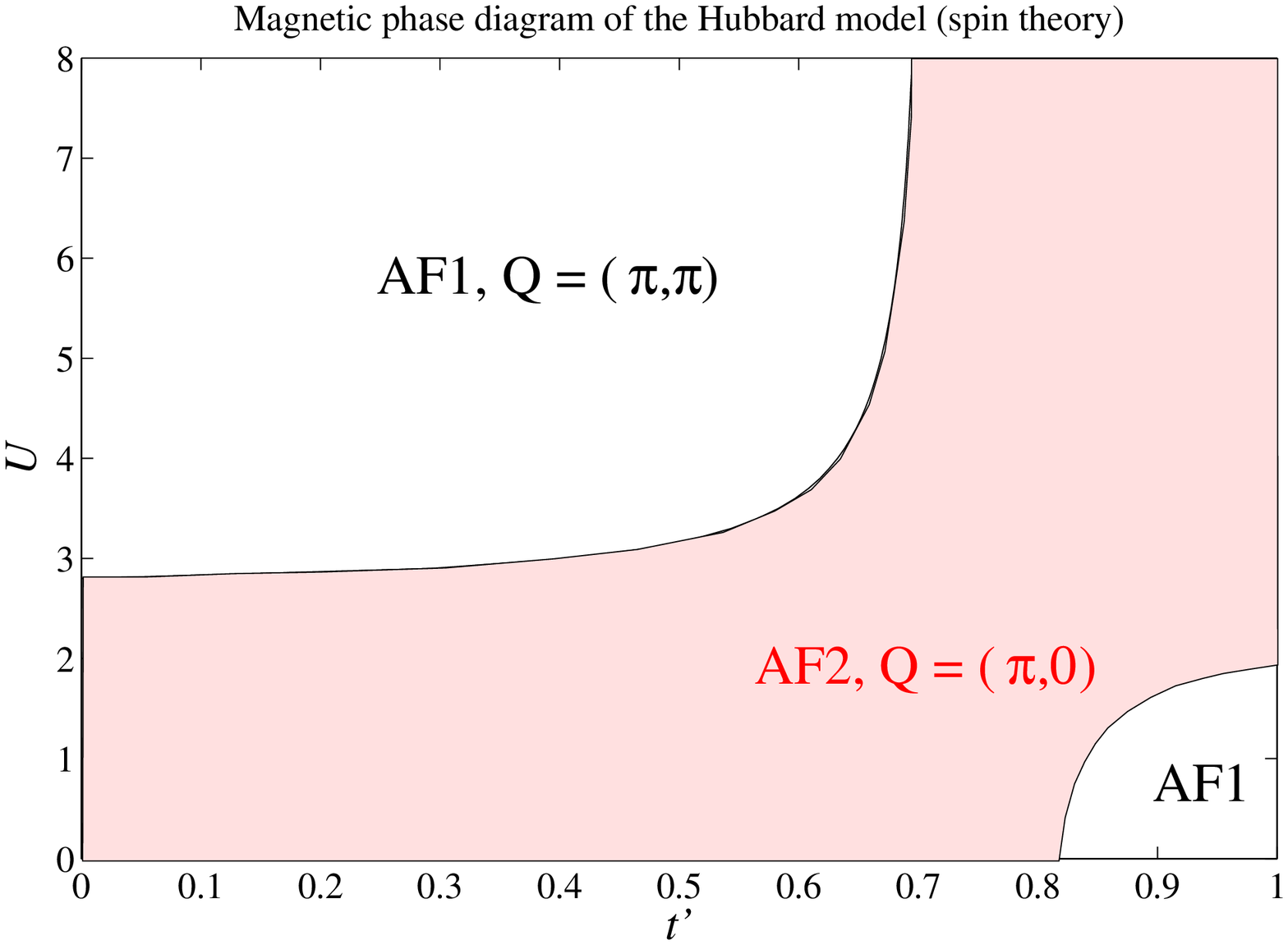}}
\end{center}
\caption
{(Color online) Classical phase diagram of the Hubbard model with
  next-nearest neighbour hopping $t'$ based on the large-$U$ expansion
  analysis. The usual AF phase (AF1) with the ordering vector
  $(\pi,\pi)$ is observed at small ratios of $t'/t$, whereas a
  so-called super-antiferromagnetic phase $(0,\pi)$ (AF2, shaded region) is realized at large
  frustration $t'/t$.
  Note that because of the nature of the
  large-$U$ expansion, this phase diagram is expected to be inaccurate
  at low $U$ values.
}
\label{Fig.classical}
\ef

Several comments on the phase diagram in Fig.~\ref{Fig.classical} are due.
First of all, the above analysis is based on
the large-$U$ expansion of the Hubbard model where electrons are localized. It should not be taken seriously
for small $U$ values where electrons can be delocalized even at half-filling. In particular, the AF2 $(\pi,0)$ phase for a large range of $t'/t$
below $U\lesssim 2.8t$ is an artifact. Secondly, the role of thermal and quantum fluctuations has
been completely neglected in the above classical analysis. The latter
are going to lower the ground state energies of the two AF phases but,
more importantly, they may favour a different type of order with no
broken rotational spin symmetry, such as the valence bond solid or the
RVB spin liquid state mentioned earlier in relation with the
$J_1$-$J_2$ Heisenberg model.

Below, we take into account the effect of short-range quantum fluctuations at
zero temperature as well as the possibility for electrons to delocalize. In order to achieve this we perform
VCA calculations of the Hubbard model with nearest and next-nearest
neighbour hopping and obtain the quantum analogue of the phase
diagram shown in Fig.~\ref{Fig.classical}, including the possibility of d-wave superconductivity.
These results are presented in Section~\ref{Sec.results}.

\section{Variational cluster approach}\label{Sec.VCA}

Despite the apparent simplicity of the Hubbard model, its phase
diagram is extremely rich in physical phenomena, such as antiferromagnetism,
incommensurate spin-density wave and d-wave superconductivity.
Analytical progress is severely hindered by the fact that the model does
not have a small parameter in the interesting regime, and, consequently, a number of numerical
methods have been proposed to treat the Hubbard model.
Among these, quantum cluster methods~\cite{Maier_RMP05} form a separate group, which
seems to successfully capture many physical features of the model,
including d-wave superconductivity that was studied~\cite{Senechal:2005,Kancharla:2005,Maier:2005,Tremblay06_review} in the context
of the cuprates.

In this work we have used the so-called variational cluster approximation (VCA),
sometimes referred to as variational cluster perturbation theory
(VCPT) in the literature. It is a special case of the self-energy
functional approximation~\cite{SFA, SFA_applied}.
The idea of this approach consists in expressing the grand canonical
potential  $\Omega$ of the model as a functional of the self-energy
$\Sigma$:
\bq
\Omega[\Sigma] = F[\Sigma] - \Tr\ln\left(-G_0^{-1}+\Sigma\right),
\label{funct}
\eq
where $G_0$ is the bare single-particle Green's function of the
problem and $F[\Sigma]$ is the Legendre transform of the
Luttinger-Ward functional, the latter being defined by the infinite
sum of vacuum skeleton diagrams.

The functional (\ref{funct}) can be proven to be stationary at the
solution of the problem, i.e. where $\Sigma$ is the
physical self-energy of the Hubbard model,
\bq
\left. \frac{\vd\Omega[\Sigma]}{\vd\Sigma}\right|_{\mbox{sol}} = 0.
\eq
The problem of finding the solution is then reduced to minimizing the
functional $\Omega[\Sigma]$ with respect to the self-energy $\Sigma$.
Two problems stand in the way however. Firstly, the functional
$F[\Sigma]$ entering Eq.~(\ref{funct}) is unknown and thus has to be
approximated in some way. And secondly, there is no easy practical way of varying
the grand potential with respect to the self-energy.

Potthoff suggested~\cite{SFA} an elegant way around these
problems by noting that since the functional $F[\Sigma]$ is a
universal functional of the interaction part of the Hamiltonian only
(i.e. the last term in Eq.~(\ref{Hubbard})), it can be obtained from the known
(numerical) solution of a \emph{simpler} reference system with the
Hamiltonian $H'$ defined on a partition of the infinite lattice into disjoint clusters,
provided that the interaction term is kept the same as in the original
Hamiltonian.
For such a cluster partition, Eq.~(\ref{funct}) can now be rewritten as
\bq
\Omega'[\Sigma] = F[\Sigma] - \Tr\ln\left(-G_0'^{-1}+\Sigma\right),
\label{funct'}
\eq
where the prime denotes the quantities defined on the cluster, to distinguish from
those of the original problem.
Combining Eqs.~(\ref{funct}) and (\ref{funct'}), we finally obtain
\bq
\Omega[\Sigma] = \Omega'[\Sigma] +
\Tr\ln\left(-G_0'^{-1}+\Sigma\right) -\Tr\ln\left(-G_0^{-1}+\Sigma\right).
\label{VCA}
\eq

Equation~(\ref{VCA}) is the central equation of the variational cluster
approximation. The role of variational variables is played by some
one-body parameters $\{h'\}$ of the cluster Hamiltonian, so that one looks for
a stationary solution
\bq
\frac{\pd\Omega}{\pd h'} \equiv
\frac{\vd\Omega[\Sigma]}{\vd\Sigma}\frac{\pd\Sigma}{\pd h'} = 0.
\label{stationary}
\eq
It has been shown by Potthoff~\cite{SFA} that the VCA and
another widely known method, cluster dynamical mean field theory
(CDMFT), can both be formulated in the framework of the above self-energy
formalism.
The particular advantage of the VCA is that it enables one to easily study
broken-symmetry phases for clusters of varying sizes. The Weiss fields $h'$ are introduced
into the cluster Hamiltonian and the potential $\Omega$ is
minimized with respect to it.
It is important to stress that, in contrast with the usual
mean-field theories, these Weiss fields are \emph{not} mean fields, in a sense
that the interaction part of the Hamiltonian is not factorized in any
way and short-range correlations are treated exactly. The Weiss fields
are introduced simply to allow for the possibility of a specific
long-range order, without ever imposing this order on the original Hamiltonian.

In this work we have defined the cluster Hamiltonian with appropriate
Weiss fields  as follows:
\begin{eqnarray}
H' = \sum\limits_{\xx,\xx',\sigma}t_{\xx \xx'} c_{\xx\sigma}^\dagger
c_{\xx'\sigma}
- \sum\limits_{\xx,\xx'}\left(\tilde{\Delta}^\dagger_{\xx\xx'}c_{\xx\up}c_{\xx'\dn} +
h.c.\right)  \notag \\
- \tilde{M}\sum\limits_{\xx,\sigma}e^{i\QQ\xx}(-1)^\sigma n_{\xx\sigma}
- \mu'\sum\limits_{\xx\sigma}n_{\xx\sigma} +
U\sum_{\xx}n_{\xx\up}n_{\xx\dn}
\label{weissfields}
\end{eqnarray}
Here as before, $c_{\mathbf{x}\sigma}^\dagger$ is the electron creation
operator at site $\mathbf{x}$ with spin $\sigma$, $n_{\mathbf{x}\sigma}$ is
the particle number operator, and $\tilde{M}$ and
$\tilde{\Delta}_{\mathbf{x}\mathbf{x}^{\prime}}$ are the Weiss fields
corresponding to the antiferromagnetic order parameter (with ordering wave-vector $Q$) and
to the superconducting order parameters respectively. For singlet
superconductivity we have $\tilde{\Delta}_{\xx\xx'}=\tilde{\Delta}_{\xx'\xx}$. In particular, for $d_{x^2-y^2}$ symmetry the Weiss field is defined as
follows ($\mathbf{e}$ is a lattice vector):
\begin{equation}
\tilde{\Delta}_{\mathbf{x},\mathbf{x}+\mathbf{e}} = \left\{%
\begin{array}{rl}
\tilde{D}, \;\; & \text{for}\;\, \mathbf{e}=\pm\hat{x} \\
-\tilde{D},\;\; & \text{for}\;\, \mathbf{e}=\pm\hat{y}.
\end{array}
\right .  \label{Delta}
\end{equation}
The corresponding order parameters, $M$ and $D$, are given by the terms
multiplying $\tilde{M}$ and $\tilde{D}$ respectively
in the Hamiltonian~(\ref{weissfields}).

In addition to the AF and SC Weiss fields, we also allow the cluster
chemical potential $\mu'$ to vary, to insure internal thermodynamic consistency~\cite{Aichhorn06} of the calculation.
Therefore, in all calculations reported in the present work, the
cluster chemical potential $\mu'$ was treated as a
variational parameter, along with symmetry-breaking Weiss fields,
such as the staggered magnetization in the AF case.

\bef[tbp]
\begin{center}{\includegraphics[width=8.6cm]{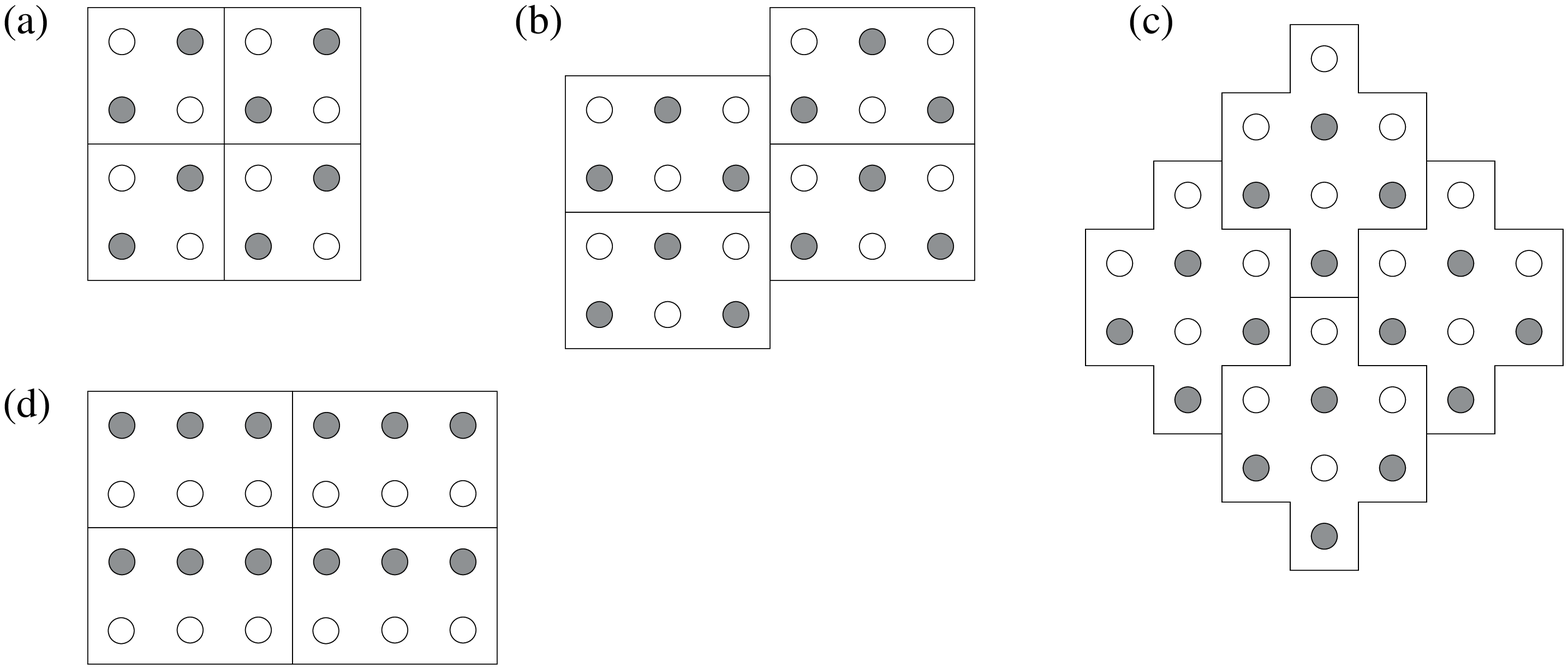}}
\end{center}
\caption
{The tilings of the square lattice with the various clusters used in
  this work containing (a) 4 sites,  (b) 6 sites, (c) 8 sites.
 In this example the grey and white sites are inequivalent since the
  $(\pi,\pi)$ AF order is possible. Note that in some cases, such as
  that of the 6-site cluster, a different tiling (d) must be chosen for a
  $(0,\pi)$ phase to become possible.
}
\label{Fig.clusters}
\ef

We solve the cluster problem using the Lanczos exact
diagonalization technique, which enables one to
find the ground state of the model at zero temperature.
The cluster Green's function $G_{ab}'(\om,\kk)$, defined for a pair of
cluster sites ($a$, $b$), was then evaluated using the so-called band Lanczos
method~\cite{BL} that offers a significant computational advantage
over other approaches~\cite{PotthoffBL}.
The search for a stationary solution Eq.~(\ref{stationary}) was
performed using a combination of the Newton-Raphson~\cite{NR} and the conjugate
gradient methods~\cite{NR}.

Since all the calculations are performed in
the grand-canonical ensemble, the requirement on the filling $\la n\ra=1$
is achieved by choosing the appropriate value of the lattice chemical
potential $\mu$. We note that this task is not trivial in
the present study since the variation of both the cluster chemical
potential $\mu'$ and
the Weiss field (the SC $\tilde{D}$ and AF $\tilde{M}$) tend to
greatly influence the value of $\la n\ra$.
The appropriate value of the lattice chemical potential $\mu$ therefore
had to be chosen at each point in the phase space of the parameters
$t$, $t'$, $U$ of the Hamiltonian~(\ref{Hubbard}) to guarantee that
the system always remained at half-filling.

In general the phase diagram will depend
on the choice of the reference cluster system $H'$ that is solved
numerically to obtain the quantities entering the VCA
equation~(\ref{VCA}).
The VCA solution becomes exact only in the thermodynamic limit of
infinitely large cluster. In practice, the typical cluster size is limited
to a maximum of 10-12 sites since the Hilbert space of the reference cluster
Hamiltonian grows exponentially with the cluster size and so does the
computational cost of the exact diagonalization algorithm.
In this work we have studied clusters of 4, 6 and 8 sites, as
depicted in Fig.~\ref{Fig.clusters}. This appears sufficient to suggest what the result should look like in the thermodynamic limit.

\section{Magnetism and Mott physics in the frustrated Hubbard model}\label{Sec.results}

\subsection{Magnetic phase diagram}\label{Sec.magnet}

\bef[tbp]
\begin{center}{\includegraphics[width=8.6cm]{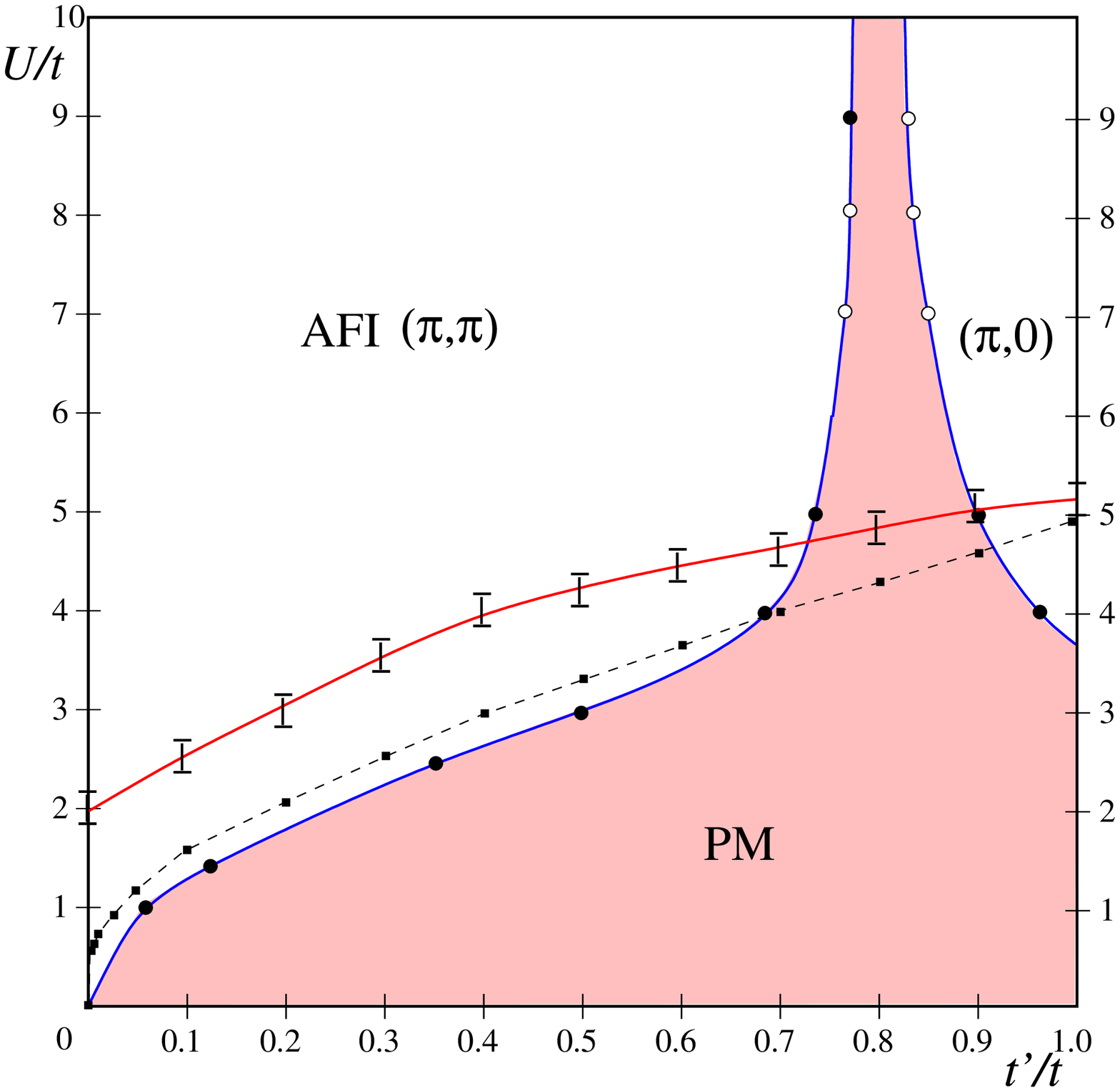}}
\end{center}
\caption
{(Color online) The magnetic phase diagram of the $t-t'-U$ Hubbard model ignoring the
  SC solution. There are two magnetic phases: ($\pi$, $\pi$),
  ($\pi$, 0) and the paramagnetic region (shaded area) where both
  order parameters vanish. The diagram was obtained with the VCA method using an
  8-site cluster. The dashed line with filled data points shows the
  Hartree-Fock prediction from Ref.~\onlinecite{Hofstetter98} for the transition between the N\'eel ($\pi$,
  $\pi$) and the non-magnetic phases. The solid line (red online) with error-bars
  indicates the transition $U_c(t')$ between the insulating
  (Mott-like) phase above and the metallic region below $U_c$ when no magnetic
  order is allowed.
}
\label{Fig.B8-AF}
\ef

When applying the VCA method to the frustrated Hubbard model, we have
studied the possibilities for both long-range magnetic order and
(d-wave) superconductivity.
In order to shed more light on the interplay between frustration and
magnetism and to connect with existing studies on spin
Hamiltonians summarized in Sec.~\ref{Sec.J1-J2}, we first report our results for purely magnetic
phases, i.e. ignoring the SC solution for the moment.
The main results of this study can be summarized by the phase
diagram in Fig.~\ref{Fig.B8-AF}
where the horizontal axis is a measure of the frustration $t'/t$ and the vertical axis the
interaction strength $U/t$.

We have looked for the same two AF phases that are predicted by both the
$J_1$--$J_2$ Heisenberg model (Sec.~\ref{Sec.J1-J2}) and the large-$U$
expansion of the Hubbard model (see Sec.~\ref{Sec.large-U}), namely
the usual N\'eel phase with the ordering wave-vector $Q=(\pi, \pi)$
and the so-called collinear order with $Q=(\pi, 0)$ (or equivalently,
$(0, \pi))$. The regions of stability of these two phases are shown on
the phase diagram in Fig.~\ref{Fig.B8-AF} for the largest cluster studied (8 sites).
The two magnetic phases are separated by a paramagnetic region
(filled area in Fig.~\ref{Fig.B8-AF}) where no
non-zero value was found for either of the two order
parameters. We shall refer to this paramagnetic region as
``disordered'' although, strictly speaking, we cannot exclude the
possibility of some other magnetic long-range order with, for example,
incommensurate wavevector $Q$, which is not tractable by our method because
of the finite cluster sizes~\cite{Hassan:2008}.

As is clear from  Fig.~\ref{Fig.B8-AF}, for large $U$ values the
disordered region is centred around the critical value
$t'/t=1/\sqrt{2}$, confirming the predictions of the large-$U$
expansion (c.f. Fig.~\ref{Fig.classical}).
This region then becomes broader as $U$ is lowered, engulfing
the whole of the phase diagram in the limit of $U=0$. This is quite
different from the semi-classical phase diagram of Fig.~\ref{Fig.classical}, that predicts that the AF2
phase should be more stable below $U\lesssim 2.8t$ for a broad range of
$t'$ values. This discrepancy is however not surprising given that the classical phase
diagram was obtained from the large-$U$ expansion that is bound
to fail in the small-$U$ region of the phase diagram.

One should note that around $t'/t=1/\sqrt{2}$ the topology of the non-interacting Fermi surface changes, as depicted in Fig.~\ref{FermiSurface}. This figure suggests why the ordering wave vectors take the above mentioned values.

\bef[tbp]
\begin{center}{\includegraphics[width=8.3cm]{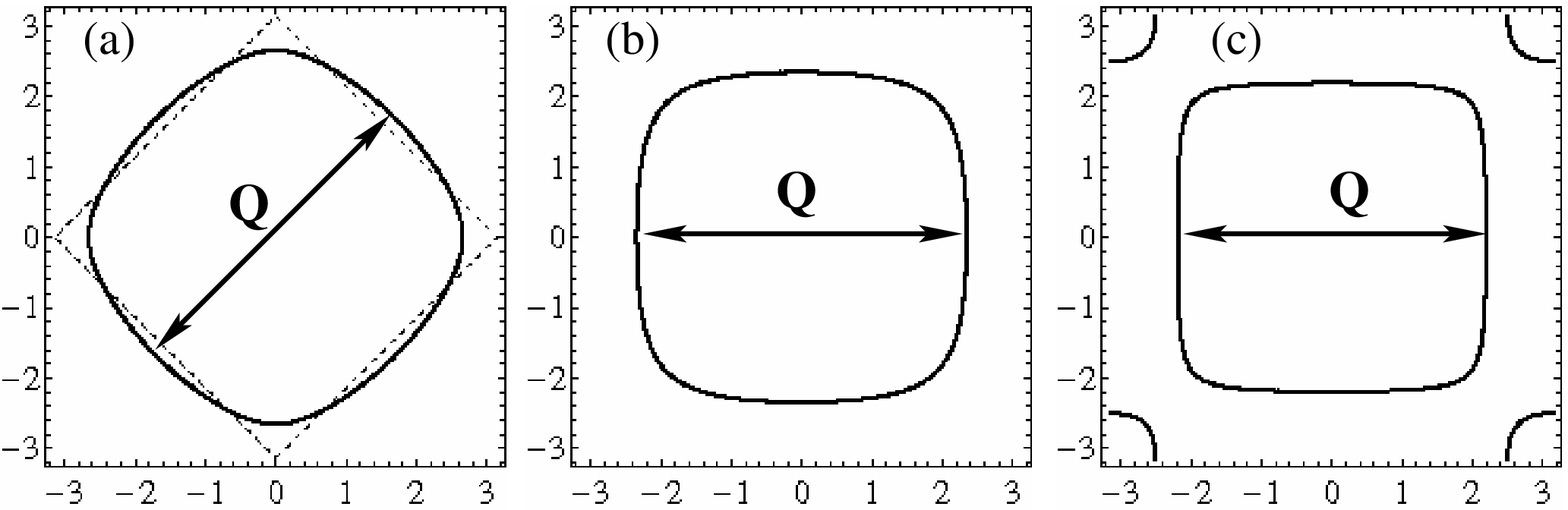}}
\end{center}
\caption{
Fermi surface for $t'=0.2t$ (left), $t'=0.6t$ (center) and $t'=0.8t$ (right). The change in topology (Lifshitz transition) occurs around $t'=0.71t$.
The best nesting vector ${\bf Q}$ is also shown. The Fermi surface for negative $t'/t$ looks like the ones above if one translates the origin to $(\pi,\pi)$. The phase diagram in other figures depends only on the absolute value of $t'/t$. }
\label{FermiSurface}
\ef

It is instructive to compare the
transition into the N\'eel phase obtained here, with the known
Hartree-Fock result for the half-filled Hubbard
model~\cite{Hofstetter98} (dashed line in Fig.~\ref{Fig.B8-AF}). The
VCA results follow closely the Hartree-Fock results for $t'<0.7$,
whereas for higher levels of frustration, the non-trivial disordered
region is revealed by VCA, followed by the $(\pi,0)$ magnetic phase
(the latter was not considered by the authors of Ref.~\onlinecite{Hofstetter98}).

We note that all the transitions on the phase diagram have been found
to be first order (with a possible exception of very low $U$ values
where a reliable solution is progressively harder to obtain). By
this we mean that the magnetic order parameter jumps as
the transition line is crossed, so that either no magnetic
solution was found in the PM phase (typical the low $U$ region) or alternatively, since hysteresis is expected,
its free energy was found to be actually higher than that of the PM solution
(typically the case for $U/t>6$).

  The two AF phases mentioned above,($\pi$,$\pi$) and ($\pi$,0) have been found
  not only in studies of the $J_1$-$J_2$
  Heisenberg model, as already mentioned, but also
  in work on the frustrated Hubbard model by Mizusaki and
  Imada~\cite{Mizusaki06}. There, the ($\pi$,0)
  phase is, perhaps confusingly, called a ``stripe'' phase. By using a
  path-integral Monte-Carlo
  technique, they found a phase diagram very similar to ours.
  In addition, these authors find a third phase with a larger
  periodicity in real space corresponding to the ordering vector
  $\mathbf{Q}=(\pi,\pi/2)$, which exists in a narrow region of
  $t'/t$ between 0.6 and 0.8 for $U\gtrsim 7$. This is precisely the
  region where neither ($\pi$,0) nor ($\pi$,$\pi$) phases have been
  found stable in this work. It may well be that another
  magnetically ordered phase (probably
  incommensurate~\cite{Hassan:2008}) exists in this
  intermediate region. Unfortunately, the clusters used in this work, see Fig.~\ref{Fig.clusters},
  are not suitable to check for the stability of the $(\pi,\pi/2)$ phase.

Apart from the possibility of some non-trivial magnetic order,
the question remains open whether the `disordered' phase around $t/t'=\sqrt{2}$ can actually
be formed by spin singlets sitting on bonds, as e.g. in the so-called
valence-bond solid (VBS). This phase is characterised by spontaneously
broken translational symmetry, but preserves the spin-rotational symmetry and
is considered to be the most likely candidate for the intermediate
phase around the boundary between the two magnetic phases
of the $J_1$-$J_2$ Heisenberg model (see Ref.\onlinecite{review_VBS} and references therein).
Another possibility is the spin liquid phase, which is similar to VBS,
but preserves the translational invariance of the system and can be
interpreted as a resonating valence bond (RVB) state.

   As it stands, the VCA approach does not permit to study the various possibilities for a
VBS or spin liquid phase. Therefore, at present, we cannot judge whether such
an order may exist in the paramagnetic region or how far it extends into
the low-$U$ part of the phase diagram.
It is interesting that a quantum spin liquid state has been recently
shown to exist in the Hubbard model on a square lattice at half
filling by numerical studies on finite clusters~\cite{Mizusaki06} using path-integral
Monte-Carlo method~\cite{PIMC}. This state, observed in a narrow
region of $U/t$ values between 4 and 7, falls between the metallic
paramagnetic state and the magnetic insulator and is believed to be
caused by charge fluctuations in the vicinity of the Mott transition (see more
on that in Sect.~\ref{Mott}). Unlike in the magnetically ordered
state, the quantum spin liquid is characterized by the absence of any
sharp peaks in the spin correlation function $S(\qq)$.
It is
certainly an intriguing possibility that should  be verified with
other existing numerical methods.

Naturally, if the VBS or quantum spin liquid solution happens to have a lower energy than
either of the two magnetic phases discussed in this work, this would
lead to further enlargement of the range of $t'$ values where no
long-range magnetic order is found. In this sense the VCA method only gives a lower bound on the
extent of the non-magnetic hatched region in  Fig.~\ref{Fig.B8-AF}, which for e.g. $U/t=9$
exists in the range of $0.77<t'/t<0.82$.
It is useful to compare these figures with the exact diagonalization results~\cite{Schulz92-96} for the
$J_1$-$J_2$ Heisenberg model discussed in Sec.~\ref{Sec.J1-J2}: There, the non-magnetic region appears in the range
\mbox{$0.4\lesssim J_2/J_1\lesssim0.6$}. For the Hubbard model, this translates into the large $U$ limit and the window $0.63\lesssim t'/t\lesssim0.78$, which is indeed
much wider than predicted by VCA.

\subsection{Metal-insulator transition}\label{Mott}

We next turn to the subject of the metal-insulator transition in
the frustrated Hubbard model at half-filling in the absence of long-range order. Since there is no bath in VCA as we define it, metallic states are less favored than in CDMFT. In the normal state, the bath present in CDMFT or DMFT can play the role of a metallic order parameter. While metallic states can occur in VCA, they cannot occur as first order transitions because of the absence of this metallic order parameter. So, contrary to the case of CDMFT~\cite{parcollet,Kyung06},
the Mott transition cannot be observed as a cusp or discontinuity in the site double
occupancy $\langle n_\up n_\dn \rangle$ -- the dependence of this
quantity on $U/t$ is a very smooth monotonic curve.

The Mott
transition is however firmly established in the half-filled Hubbard
model, and can indeed be observed by analyzing the spectral function  $A(\kk,\omega)$.
Therefore, for the purpose of this
study, we define a ``metal'' as a state with non-vanishing spectral
function at the Fermi level, $A(\kk,\omega=0)$. We note in passing that the latter definition is actually broader than saying that there exists a
well-defined Fermi surface in the ground state, for the following
reasons.
Firstly, the regions of non-vanishing  $A(\kk,\omega=0)$ need not  form a
closed surface, but may instead have a shape of isolated arcs (c.f.
the well-known Fermi arcs as revealed by ARPES spectroscopy~\cite{ARPES}
in the underdoped cuprates).
Secondly, the Landau picture of a metal predicts an infinite lifetime
for the quasiparticles at the Fermi surface, equivalent to
the requirement of delta-function shape for $A(\kk,\omega=0)$ at the Fermi energy.

Since the presence of a long-range magnetic order opens up a gap at the
Fermi surface, we intentionally suppress the
possibility of magnetic ordering. Only the effect of short-range magnetic correlations is included in the cluster. This is an established practice used to obtain the
parameters of the metal-insulator transition~\cite{parcollet,Maier_RMP05,Tremblay06_review,Kyung06}.

The procedure we have adopted is as follows. For a  given value of $t'$,
we plotted the function $A(\kk,\omega=0)$ across the Brillouin zone (BZ) for
several values of the interaction $U$, with the Lorentzian broadening
$\eta=0.05$ used to account for the imaginary part of $A(\kk,\omega)$. The point where the
spectral function vanishes everywhere in the BZ (as $U$ is increased)
marks the transition from metallic to
insulating state. In the present approach, the transition appears as a crossover, that strictly speaking at zero temperature should be a second-order transition. On the anisotropic triangular lattice, it was found that as a function of $t'$, the Mott transition goes from second order at small $t'$ to strongly first order at large $t'$ through a tricritical point~\cite{Kyung_private,Kyung06}.

Our ``crossover'' transition line together with error bars is plotted in
Fig.~\ref{Fig.B8-AF}. We see that with increasing frustration $t'$,
the critical value $U_c(t'/t)$ increases monotonically.
We note an important difference between the low-$t'$ region and that
for $t'\gtrsim 0.7t$. In the former case of almost perfect nesting, the
effect of short-range correlations is strong,
leading to a surprisingly low value of $U_c\approx 2t$. At first sight,
this is too different from the well-known DMFT result~\cite{DMFT} for the Mott metal-insulator
transition, $U\approx 12t$. It must be noted
however that the single-site DMFT approach~\cite{DMFT} does not take into account short-range magnetic correlations, as opposed to
cluster methods such as CDMFT or VCA. In addition, to study the Mott transition, DMFT assumes the presence of large
frustration to prevent magnetic long-range order. Hence, the DMFT result should be compared with the region $t'\gtrsim0.7t$ in our phase diagram Fig.~\ref{Fig.B8-AF} where magnetic order is naturally absent and the critical interaction strength is $U_c\approx 5t$. In real experiments, that is where we believe true Mott physics would be observed. In any case, it is physically expected that in two dimensions the critical $U$ for the Mott transition depends on frustration, as pointed out in CDMFT study of the anisotropic triangular lattice~\cite{Kyung06}.

It is encouraging that values of $U_c(t'/t)$ very similar to ours have been found in the path-integral Monte-Carlo study~\cite{Mizusaki06} of the same model,
yielding $U_c=3t$ at $t'/t=0.25$ and $U_c=5t$ for $t'/t=0.8$.
However, the authors of Ref.~\onlinecite{Mizusaki06} have not
excluded the possibility of long-range magnetic order, which is
why they observe the Mott transition happening at infinitesimally
small values of $U$ in the case of perfect nesting $t'=0$. We stress
that this is in complete agreement with our data, although we
interpret this as an opening of the AF gap at the Fermi surface rather
than Mott transition into a phase with no broken spin-rotational
symmetry. A somewhat poorer agreement is seen with the results of the optimized variational
Monte Carlo (VMC) method in Ref.~\onlinecite{Yokoyama06}. There,
the authors obtain the value $U_c\approx 7t$ in the region
$|t'/t|<0.5$ that they studied.

Our value of $U_c\lesssim 5t$ around $t'/t=0.7$ should also be compared
with the recent CDMFT results
for strongly frustrated lattices, such as
the triangular lattice~\cite{Kyung_triang07} with $U_c=10.5t$ and the asymmetric square
lattice~\cite{Kyung06} ($t'=t$ along only one diagonal) with $U_c\approx 8t$.
Although CDMFT predicts higher $U$ values for the Mott metal-insulator
transition than VCA in these cases, they clearly fall into the same
ballpark. However, for the square lattice without frustrations
($t'=0$) the 4-site cluster CDMFT gives a value~\cite{Kyung_private}
for the Mott transition $U_c\approx 5t$, quite a bit larger than our
result $U_c\approx 2t$.  Similar to CDMFT, a value $U_c\approx 6t$
was also found in Quantum Monte Carlo studies~\cite{Vekic:1993} at $t'=0$.
  This discrepancy is however not surprising since it is well
known~\cite{Kyung_private}, for the reasons mentioned at the beginning
of this section, that VCA method (without a bath) tends to
overestimate the effect of interactions compared with CDMFT, thereby yielding
smaller values of $U_c$ for the Mott transition.

\section{Interplay between magnetism and superconductivity}\label{Sec.AF-SC}
\subsection{Results and discussion}
\bef[tbp]
\begin{center}{\includegraphics[width=8.6cm]{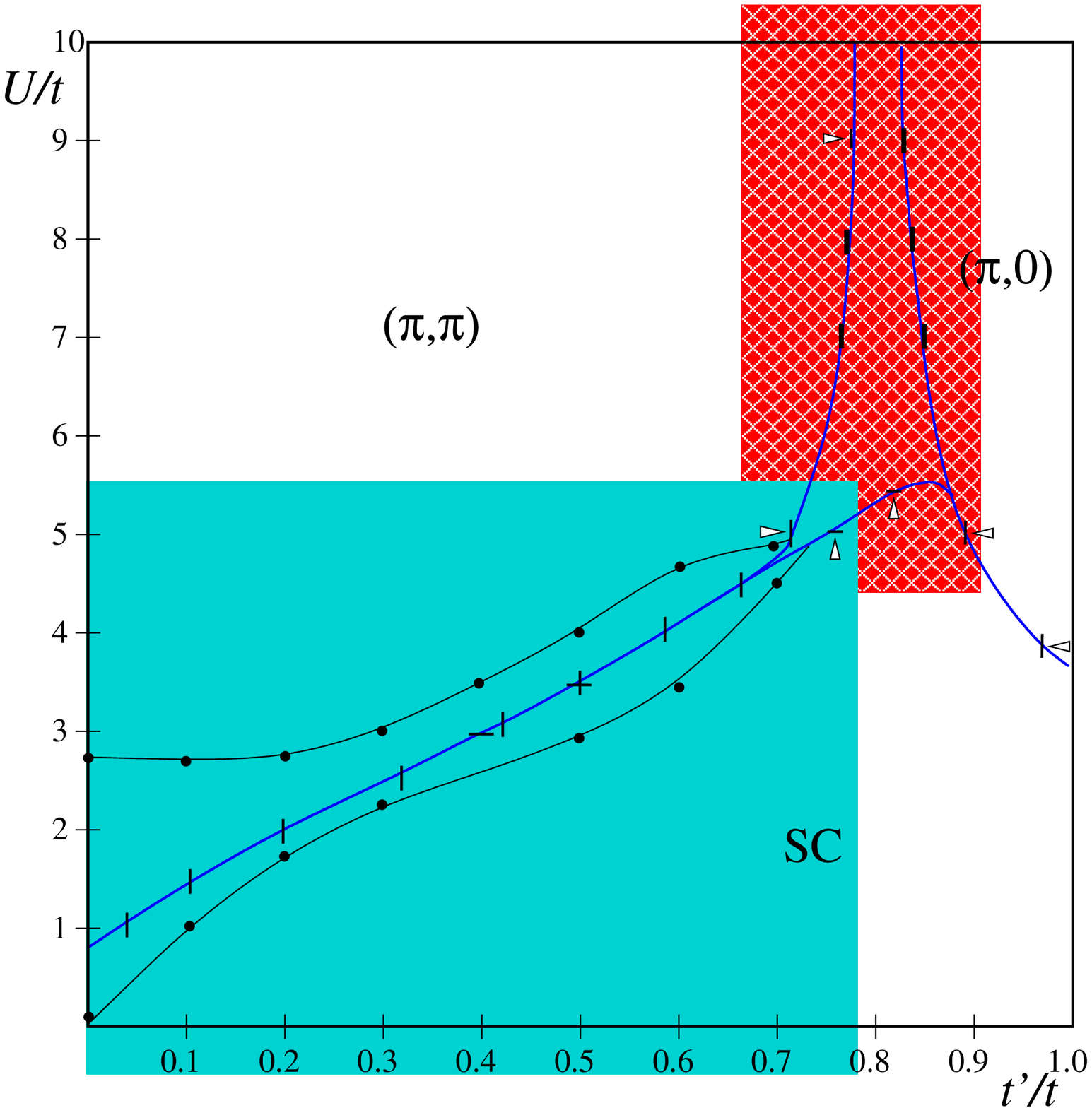}}
\end{center}
\caption
{(Color online) The phase diagram of the $t-t'-U$ Hubbard model obtained with the VCA
  based on a 8-site cluster.
  The solid lines denote the phase
  boundaries between the $(\pi,\pi)$ and $(\pi,0)$  antiferromagnetic
  phases and the d$_{x^2-y^2}$ SC phase. The hatched area on the phase
  diagram (red-and-white online)
  denotes the  critical region where neither $(\pi,\pi)$ nor $(\pi,0)$
  order could be found. The coexistence region between AF and SC
  phases (shaded area, cyan online) is contained between two black lines
  that meet around $t'=0.7t$, with the (blue) line in between indicating the
  area where the free energies of the would-be separate SC and AF phases become
  equal (c.f. the point $U=2.45t$ in Fig.~\ref{Fig.af-dx1}b).
  Triangles and filled circles denote points on the phase boundary where an order
  parameters sustains a discontinuity at a first-order phase transition;
  short dashes inside the coexistence region mark the points where
  total energies of the AF and SC phases are equal.
}
\label{Fig.B8}
\ef

\bef[tbp]
\begin{center}{\includegraphics[width=8.6cm]{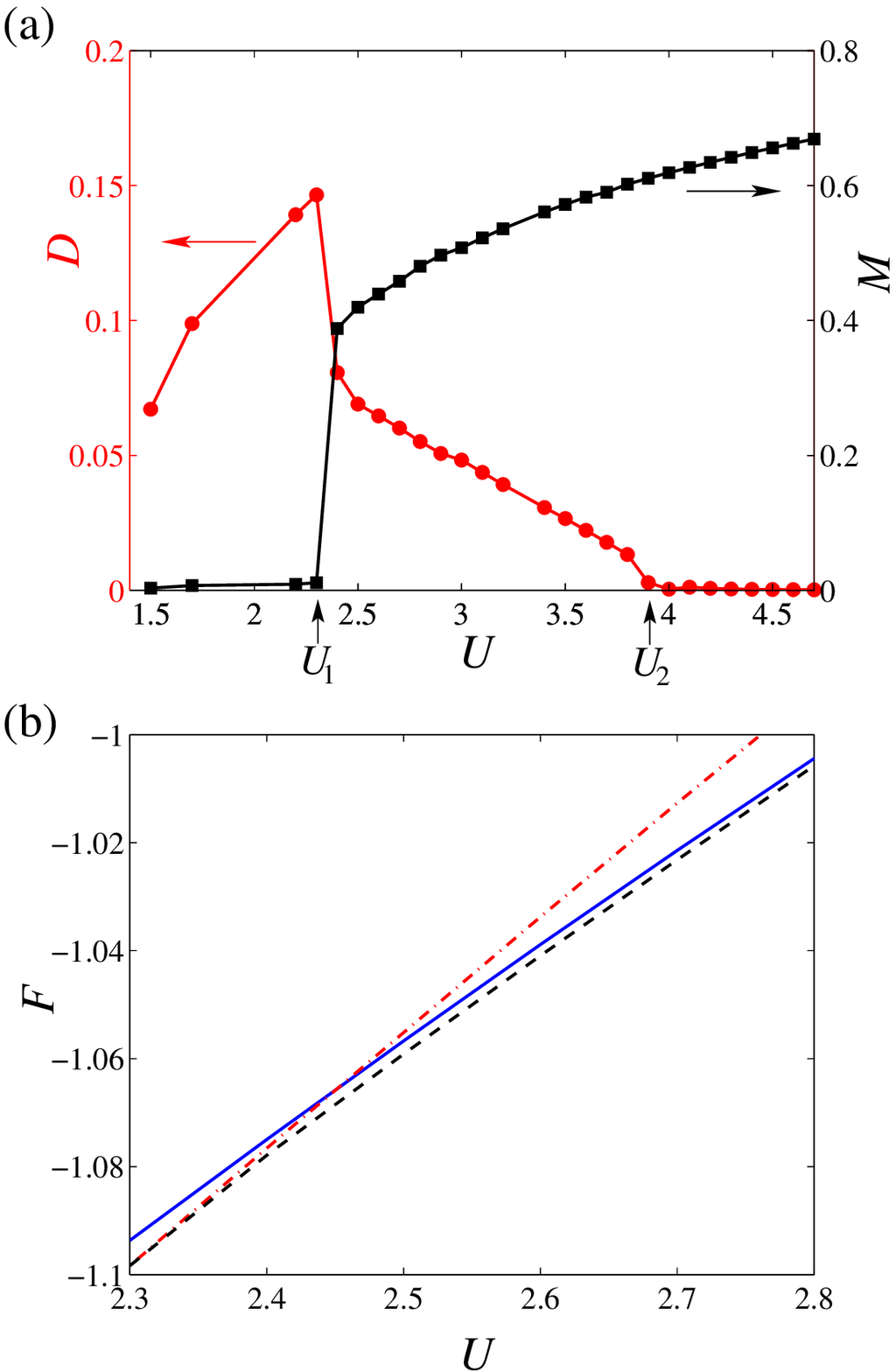}}
\end{center}
\caption{(Color online) Details of the AF and SC phase coexistence for a 2x3 cluster
  at $t'=0.2t$. (a) Expectation values of SC ($D$, circles and red
  line) and AF ($M$, squares and black line) order
  parameters as a function of increasing interaction $U$;  note
  different scales for the two quantities plotted. Values $U_1$ and
  $U_2$ denote the positions of the first-order phase transitions
  into/from the coexistence phase. (b) Blow up  of the
  free energy  $F=\Omega + \mu \langle n\rangle$ as a function of $U$ near $U=2.5t$
  is shown for all three phases studied: AF phase (blue solid line), SC
  phase (red dash-dotted line) and for the coexistence phase where
  both $D$ and $M$ order parameters are non-zero (dashed black
  line). The latter phase has lower energy than the other two in the
  whole coexistence region of $2.3 < U/t < 3.9$.
}
\label{Fig.af-dx1}
\ef

\bef[tbp]
\begin{center}{\includegraphics[width=8.6cm]{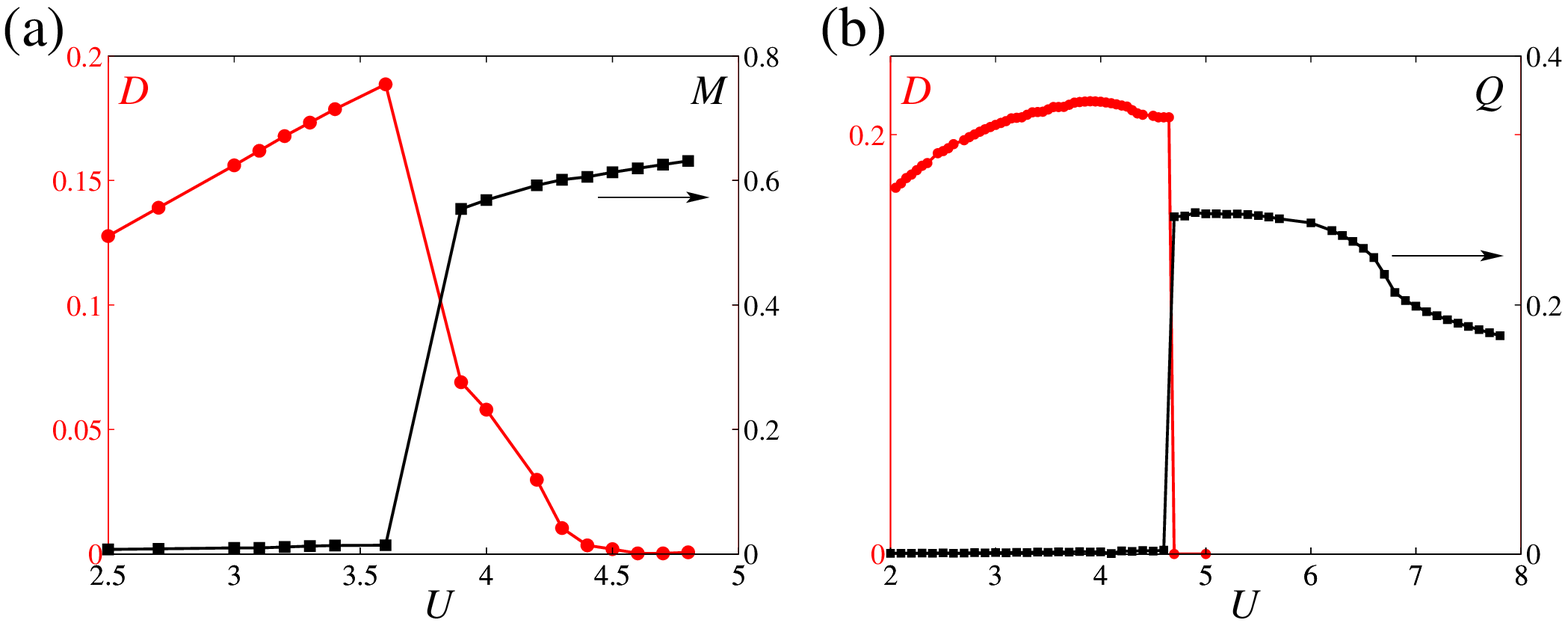}}
\end{center}
\caption{(Color online) The expectation values of SC ($D$, circles and
  red line, left-hand scale) and two types of AF order parameters (shown with squares and black
  lines),  calculated on a 2x3 cluster for
 (a) $t'=0.5t$, the AF $(\pi,\pi)$ phase order parameter $M$ (right-hand
  scale), and (b) $t'=0.8t$, the $(\pi,0)$ phase  order parameter $Q$ (right-hand
  scale). Unlike in (a), no coexistence region is seen in case (b), where the
  first-order transition occurs at $U_c=4.7t$.
}
\label{Fig.af-dx2}
\ef

Our final VCA phase diagram
is shown in  Fig.~\ref{Fig.B8} for the largest 8-site cluster studied. Most interestingly, in addition to the two magnetic phases
discussed in the previous section, a d-wave superconducting solution
(with $d_{x^2-y^2}$ symmetry) comes out
naturally from the VCA calculations for low values of $U/t$ in the phase diagram.
It is clear from Fig.~\ref{Fig.B8} that the frustration tends
to destroy the N\'eel phase and stabilise the SC solution.  In the
whole range $|t'/t|<1$, we did not
find any superconducting regions with stable $d_{xy}$ symmetry of the order
parameter, although there are indications~\cite{Hassan:2008} that such a phase would become stable in the case
of (unrealistically large) frustration strength $t'/t\gtrsim 1.1$.

In the low-$U$ region of the phase diagram, both AF and d-wave SC
solutions are stable, hence the one with lowest free energy $F$ would
win. Since we work at a fixed particle density $n$ (half-filling), we perform the
Legendre transform to obtain the free energy from the grand-canonical
potential $\Omega$:
\begin{equation}
F \equiv E-T S = \Omega + \mu \la n\ra.
\end{equation}
Moreover, since the calculations are done at zero temperature, the Gibbs free
energy in this case is identical to the total energy $E$ of the system.
We find that the transition between the magnetic and the SC phases is first
order, with the blue line in  Fig.~\ref{Fig.B8} (in the centre of the shaded region)
denoting the points where the total energies of the two competing
phases (no coexistence allowed) become equal.
The free energy of both phases is illustrated in more detail by the dot-dashed and solid lines in Fig.~\ref{Fig.af-dx1}b.

Most interestingly, the lowest energy solution (dashed line in
Fig.~\ref{Fig.af-dx1}b) corresponds to a new phase, shown as shaded area in
Fig.~\ref{Fig.B8}, where both magnetic and SC order parameters are
non-zero. This is a phase with a true {\it homogeneous coexistence} of the
magnetic and superconducting phases, which one may want to call an
\emph{antiferromagnetic superconductor} to emphasize the difference from a
more usual inhomogeneous coexistence observed e.g., at a first-order
transition.

The details of the transition between the coexistence
phase and the pure AF and SC phases are illustrated in
Figs.~\ref{Fig.af-dx1}a and  Fig.~\ref{Fig.af-dx2}a that show the
dependence of the corresponding order parameters, $M$ and $D$, on the
interaction strength $U$.
As $U$ decreases below $U_1$ in Fig.~\ref{Fig.af-dx1}a, we first
observe a first-order transition from the
coexistence phase into a pure d-wave SC state, where the AF order parameter
plunges to zero and the SC order parameter sustains an upward jump as
the coexistence phase ceases to exist.
As the interaction increases above $U=U_2$, there is a similar transition from
the coexistence phase into the pure antiferromagnet ($\pi$,$\pi$), although
this time the transition appears more continuous (see Fig.~\ref{Fig.af-dx1}a).

Clearly, our phase diagram Fig.~\ref{Fig.B8} shows that frustration $t'/t$ favors
the SC phase as long as it is not too large. Indeed SC becomes more stable and occupies a
broader region of the phase diagram as frustration increases
at low to  intermediate interaction strength $U$, until $t'/t$ becomes
large enough for the AF2 $(\pi,0)$ phase to decrease the area occupied
by the SC phase.
The latter transition is of first order and is accompanied by a sharp jump in the
values of the respective order parameters at $U=U_c$, as one can
verify on Fig.~\ref{Fig.af-dx2}b for $t'/t=0.8$ for the 6 site
cluster.
 Sometimes a very narrow hysteresis region ($\Delta
U/t \lesssim 0.2$) has been observed, depending on whether the transition is approached from above or
below the critical value of $U_c$.

In the region of $t'/t\approx 0.75$ of the phase diagram Fig.~\ref{Fig.B8}, the SC phase
has a direct boundary with the disordered non-magnetic phase
as a function of increasing $U$. Given the possibility for the
existence of the VBS order in that region, this opens up the interesting possibility of a direct transition from VBS into the SC
state. Since the two phases have symmetry groups that are not related
to each other, according to Landau theory they would be separated
either by a first order transition or by a coexistence phase. Or,
beyond the Landau paradigm, they could lead to an example of
deconfined quantum criticality~\cite{Senthil:2004}. A similar
deconfined critical transition could also possibly occur between the
VBS and antiferromagnetic phases.

\bef[tbp]
\begin{center}{\includegraphics[width=8.6cm]{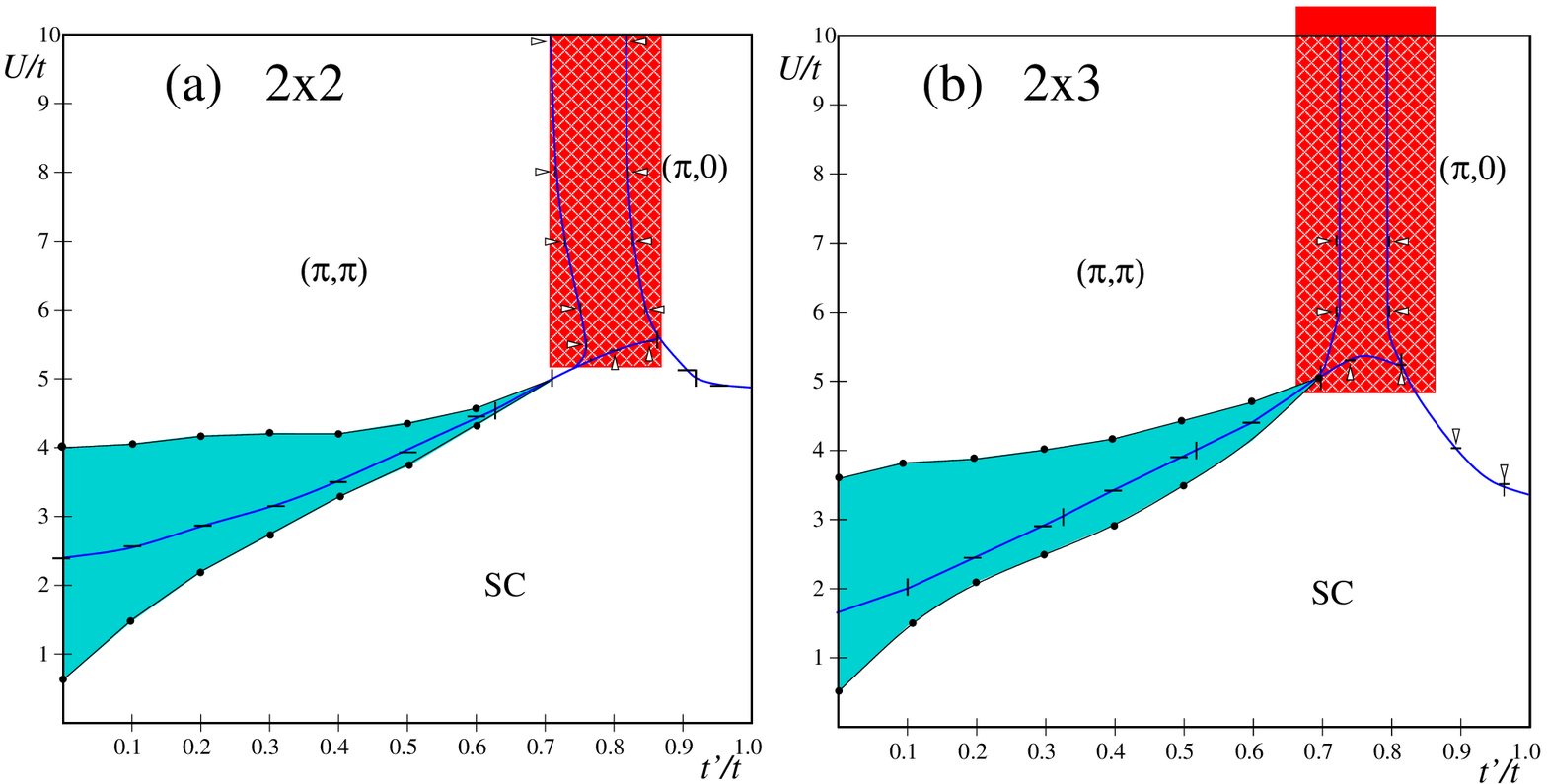}}
\end{center}
\caption
{(Color online) The phase diagram of the $t-t'-U$ Hubbard model obtained with the VCA
  based on an (a) 4-site cluster and (b) 6-site cluster.
  The hatched area (red-and-white online) denotes the  critical region where neither $(\pi,\pi)$ nor $(\pi,0)$
  order could be found or when the paramagnetic solution had lower energy then
  any of the AF phases. The coexistence region between SC anf AF
  phases (shaded area, cyan online) is shown.  The solid lines, as well as
  open triangles and short dashes on the phase boundaries have the same
  meaning as in Fig.~\ref{Fig.B8}.
 }
\label{Fig.2x2}
\ef

To assess convergence towards the thermodynamic limit, it is useful to compare the phase diagrams obtained for different
cluster sizes. Figure~\ref{Fig.2x2} shows the VCA
phase diagrams obtained from the 4-site and 6-site clusters.
We see that the main conclusions, namely that the d-wave SC phase
is the ground state for low $U$ values,  remains unchanged. However the
position of the AF1-SC phase boundary shifts to lower $U$ values with
increasing cluster size. This is to be expected since perfect nesting at $t'=0$ and half-filling should lead to antiferromagnetism as the true ground
state of the Hubbard model at arbitrary small values of $U$. This suggests that perhaps the coexistence of SC and AF1 at small $t'=0$ should disappear altogether in
the thermodynamic limit of infinitely large cluster size. We do see from Figs~\ref{Fig.2x2} and~\ref{Fig.B8} a decrease in the size of the coexistence region as the cluster size increases.

We also note that the phase boundary of the collinear $(\pi,0)$ phase occurs at lower
values of $U$ with  increasing cluster sizes. For example, comparing
Figs.~\ref{Fig.2x2} and \ref{Fig.B8}, we see that
at $t'/t=0.96$ the transition from the AF2 phase into the SC phase occurs
at $U/t\approx 3.5$ for the 8-site cluster  instead of $U/t\approx 5$
that one observes for the smallest cluster studied.
As in the case of the largest cluster studied (phase diagram in
Fig.~\ref{Fig.B8} above), all transitions between the different phases, including the
coexistence phase between AF and SC orders, were found to be first
order.

We now compare our phase diagram in
Fig.~\ref{Fig.B8} with that found by other methods. Recently, the
path-integral Monte Carlo study of the same model by Mizusaki and Imada~\cite{Mizusaki06}
has revealed a phase diagram where the magnetic and PM regions are in
very good agreement with our Fig.~\ref{Fig.B8-AF}, apart from an extra
magnetic phase that the authors of Ref.~\onlinecite{Mizusaki06} observe
between the AF and the $(\pi,0)$ phases. This however does not
contradict our results since even more complicated phases, such as
incommensurate magnetic order, may be possible but are beyond reach of
quantum cluster methods such as VCA.
Another important difference is the existence of the quantum spin
liquid phase found in Ref.~\onlinecite{Mizusaki06} that we commented upon in
Sect.~\ref{Sec.magnet}.
The possibility of superconducting phase has not been addressed however
in Ref.~\onlinecite{Mizusaki06}. This would have been very instructive in
light of our findings.

Another recent study of the half-filled Hubbard model has been
performed recently by Yokoyama{\it~et~al.}~\cite{Yokoyama06} using optimization
VMC. They considered both the $(\pi,\pi)$ AF phase and
$d_{x^2-y^2}$ superconductivity. It is puzzling that the AF phase has been
found to be limited to the region $|t'/t|<0.2$ only, in contradiction
with both our work and the known results of the $J_1$-$J_2$ model in
the strong coupling limit. As the authors themselves suggest, this
discrepancy is most likely due to a bad choice of the variational
state and/or the limitations of the VMC method. In that work, the d-wave SC state is found to be most
stable in the vicinity of the Mott transition ($U_c=6.7t$) for a narrow
range of $0.2<|t'/t|<0.4$, although the authors mention that a small
magnitude of the SC gap survives often to very small values of $U/t$,
which would be in agreement with our results showing that SC
exists in the whole region of $0<|t'/t|<1$ down to
$U=0$. Unfortunately, the limited range of  $|t'/t|<0.5$ studied in
Ref.~\onlinecite{Yokoyama06} does not help to shed light on the existence of
other AF orderings, such as the $(\pi,0)$ state, or indeed on the long
sought after quantum spin liquid state, claimed to have been observed
convincingly in Ref.~\onlinecite{Mizusaki06}.

\subsection{Comparison with results on the anisotropic triangular lattice}

To conclude this section, we contrast the results with those obtained on the anisotropic triangular lattice~\cite{Kyung06,Sahebsara06}. First of all, on that lattice
CDMFT shows that phase transitions between ordered phases occur at the same location as the Mott transition when the latter is first order~\cite{Kyung06}. This does not happen when the transition is second order~\cite{Kyung_private}. In our case, the Mott transition is always second order and the transitions between ordered phases do not coincide with the Mott line as we can see by comparing the line with error bars on Fig.~\ref{Fig.B8-AF} with the phase diagram in Fig.~\ref{Fig.B8}.

On the anisotropic triangular lattice, the transition between d-wave superconductivity and antiferromagnetism is always first order with no coexistence region, in contrast with our case where $(\pi,\pi)$ antiferromagnetism is separated from d-wave superconductivity by a coexistence phase. At larger frustration however, d-wave superconductivity is separated from $(\pi,0)$ antiferromagnetism by a first order transition. Since the triangular lattice has geometric frustration, not only frustration induced by interactions, one can speculate that it is the larger frustration on the anisotropic triangular that leads to the disappearance of the coexistence phase. However, at $t'=0$ both problems become identical and there is a clear disagreement between the results that can come only from differences in the two calculational approaches, CDMFT vs VCA.

It is rather striking that on the anisotropic triangular lattice, the d-wave order parameter is largest as a function of $U/t$ when it touches the first order boundary with the antiferromagnetic phase. As can be seen from Figs.~\ref{Fig.af-dx1}a and ~\ref{Fig.af-dx2}a, this also occurs in our case when d-wave superconductivity touches the homogeneous-coexistence phase boundary. Also, the maximum value of the order parameter increases in going from
$t'=0.2t$ to $t'=0.5t$, as in the case of the anisotropic triangular lattice. At larger frustration, the trend as a function of $t'/t$ reverses on the latter lattice. In our case, we see that at large frustration $t'=0.8t$
on Fig.~\ref{Fig.af-dx2}b, the d-wave order parameter reaches its maximum value before it hits the first order boundary with the $(\pi,0)$ phase.

It would clearly be interesting to compare what are the predictions for our case of other quantum cluster approaches, such as CDMFT and DCA. The weak to intermediate coupling Two-Particle Self-Consistent Approach~\cite{Hassan:2008} suggests that at small values of $U$ superconductivity disappears in favor of a metallic phase and that at $t^\prime=0$ antiferromagnetism dominates. We expect the predictions of VCA to be more reliable at strong coupling.

\section{Summary and conclusions}\label{Sec.conclusions}

We have analyzed the phase diagram of the half-filled Hubbard model as a
function of frustration $t'/t$ and interaction $U/t$
using both analytical and numerical techniques.

The classical analysis based on the resulting large-$U$ effective
spin Hamiltonian allowed us to draw the classical magnetic phase
diagram shown in Fig.~\ref{Fig.classical}. Of course, this classical
approach remains completely oblivious to the  role of
quantum fluctuations and in addition is designed for the high-$U$ segment of the phase diagram where electrons are localized.

To treat quantum fluctuations as well as the possibility of delocalization, we used the Variational Cluster Approximation. Because of the nature of the
VCA method, it relies on the choice of a finite (necessarily small) cluster
on which the problem can be solved exactly. In order to the finite-size
effects, we have analyzed the results for clusters of 4, 6
and 8 sites (the latter being almost at the limit of what can be
achieved with today's powerful supercomputers using the exact
diagonalization algorithm).
Although the exponentially increasing computational cost
did not permit us to study larger clusters, the apparent similarities
between the 6- and 8-site cluster solutions allow us to  conclude
that the VCA calculations reported in this study are
close to convergence with respect to increasing cluster size.

Independent of the cluster size used, the key features of the resulting
phase diagram are as follows. At large values of $U$ the VCA
results agree qualitatively with the
classical large-$U$ expansion (Sec.~\ref{Sec.large-U}) of the Hubbard Hamiltonian
and with the known results for the related $J_1$-$J_2$ Heisenberg
spin model (Sec.~\ref{Sec.J1-J2}), which show that two competing AF phases
AF1 and AF2 with ordering wave-vectors $\QQ_1=(\pi,\pi)$ and  $\QQ_2=(\pi,0)$
exist for frustrations lower and higher, respectively, than some
critical value given roughly by $t_c'/t=1/\sqrt{2}\approx 0.71$. This
is where the Fermi surface changes topology in the non-interacting
case (see Fig.~\ref{FermiSurface}).
These two magnetic phases are separated by a disordered region where
we don't find non-zero values of either order parameter.
The Heisenberg model studies point to possible existence of an
exotic valence bond solid (VBS) phase around the critical frustration
value $t'_c$. Although the direct study of the VBS phase remains beyond reach
of quantum cluster methods such as VCA, it is
encouraging that the obtained phase diagram exhibits a region around
the critical frustration $t_c$ where no long-range AF order could be found.

We have also addressed the issue of the role of frustration on the
metal-insulator transition that is known to exist in the Hubbard model
at half-filling. The distinction between metallic and insulating
phases was based on the analysis of the spectral function
$A(\kk,\omega)$ at the Fermi level, which is vanishing in the
insulating ground state. We find that the value of the
interaction strength $U_c$ where the insulator appears
rises monotonically as a function of frustration strength
$t'$. The value of $U_c$ turns out to be surprisingly low for small
$t'$ values ($U_c\approx 2t$). Low values are expected
from the effect of short-range antiferromagnetic fluctuations
that are particularly strong near perfect nesting at
$t'=0$. Nevertheless, comparisons with other results definitely
suggest that VCA overemphasizes the effect of $U$ so that the
insulator-metal transition at half-filling should be closer to the
CDMFT and QMC values $U_c\approx 5-6t$.
For large frustration near $t_c'/t=1/\sqrt{2}\approx 0.71$, the
transition occurs at larger values of $U_c\approx 5t$. Since neither
the AF1 or AF2 phases are stable in this highly
frustrated region, this is where one would experimentally be more
likely to see a genuine Mott transition at finite temperature where
ordered phases are absent.

Most importantly, the VCA method allowed us to study the
region of the phase diagram with low to intermediate interaction
strength, which is inaccessible in the large-$U$
expansion or the $J_1$-$J_2$ Heisenberg spin model.
We find that even at half-filling, where the
tendency towards the antiferromagnetic ordering is strong, frustration
allows d-wave superconductivity to appear for a range of values of
$U/t$ that generally increases with frustration since the latter is
detrimental to $(\pi,\pi)$ antiferromagnetism. With frustration in the
range $t_c'/t\approx 1/\sqrt{2}$, both
$(\pi,\pi)$ and $(\pi,0)$ antiferromagnetism disappear but d-wave
superconductivity survives for $U_c \lesssim 5t$. Increasing
frustration further favors the $(\pi,0)$ phase but d-wave
superconductivity continues to appear at  smaller values of $U/t$. All
the phase transitions are first order, except possibly the transition
from the coexistence phase into the $(\pi,\pi)$ magnet, which is weakly first
order (note a very small jump in the value of SC order parameter at
the $U=U_2$ phase boundary in Fig.~\ref{Fig.af-dx1}~a).
In the case of $(\pi,\pi)$ antiferromagnetism, the transition to pure d-wave superconductivity
occurs through a region where both phases coexist homogeneously. Finite size analysis suggests that this
coexistence region is relatively robust, although its boundaries
shrink with increasing cluster size. Coexistence may disappear in the
thermodynamic limit.

An important prediction of our study for experiments is that d-wave
superconductivity may appear by applying sufficiently high pressure on
the half-filled parent compounds of high-temperature superconductors. This type
of transition is observed in layered BEDT organics~\cite{BEDT_Cl} and
can be explained by the Hubbard
model~\cite{Sahebsara06,Kyung06}. Hence positive results of such an
experiment on the cuprates would spectacularly help to establish definitively the electronic origin of d-wave superconductivity.

It would be interesting to pursue the issues addressed in this work with other quantum cluster approaches and also to study
the case of doped Hubbard model away from half-filling, with possible
comparison with the results of the much studied $t$-$J$
model. Anticipating on the results, it should be easier to reach the
d-wave superconducting state by applying pressure on a slightly doped
insulating parent than on the half-filled insulator. These
issues, of much relevance to the physics of high-temperature
superconductivity in the cuprates, are left for future studies.

\section*{Acknowledgements}
The authors are grateful to Raghib Syed Hassan, Bumsoo Kyung, and Bahman
Davoudi for many stimulating discussions.
A.H.N. was supported by FQRNT.

\noindent 
C.S. acknowledges partial support by the FWF (Project no. P18551-N16) and by KUWI 
grant of Graz University of Technology. 
Numerical computations were performed on  the Dell clusters of the
R\'eseau qu\'eb\'equois de calcul de haute performance (RQCHP) and on
Sherbrooke's Elix cluster.
The present  work was supported by NSERC (Canada), FQRNT (Qu\'{e}bec),
CFI (Canada), CIFAR, and the Tier I Canada Research chair Program (A.-M.S.T.).



\end{document}